\documentclass[preprint,12pt]{elsarticle}

\usepackage{amsmath}
\usepackage{ulem}

\newcommand{\mrm}[1]{\mathrm{#1}}

\newcommand{\kC}[0]{k_\mrm{C}}

\newcommand{\Zc}[0]{Z_\mrm{c}}

\def\d{\mathrm{d}}

\def\QQ{\mathbf{Q}}
\def\pp{\mathbf{p}}

\def\kk{\mathbf{k}}

\def\rr{\mathbf{r}}
\newcommand{\mR}[0]{m_\mrm{R}}
\newcommand{\Mtot}[0]{M_\mrm{tot}}

\newcommand{\Stot}[0]{S_\mrm{tot}}

\usepackage{color} 
  
  \def\nuc#1#2{\relax\ifmmode{}^{#1}{\protect\text{#2}}\else${}^{#1}$#2\fi}
  \def\itnuc#1#2{\setbox\@tempboxa=\hbox{\scriptsize\it #1}
    \def\@tempa{{}^{\box\@tempboxa}\!\protect\text{\it #2}}\relax
    \ifmmode \@tempa \else $\@tempa$\fi}

\usepackage{amssymb}

\journal{Annals of Physics}

\begin{document}

\begin{frontmatter}


\title{Range corrections in Proton Halo Nuclei}


\author[chalmers]{Emil Ryberg}
\author[chalmers,utk,ornl]{Christian Forss\'en}
\author[darmstadt,emmi]{H.-W. Hammer}
\author[chalmers,utk,ornl,anl]{Lucas Platter\corref{lp}}
\cortext[lp]{lplatter@utk.edu}

\address[chalmers]{Department of Fundamental Physics, Chalmers
  University of Technology,\\ SE-412 96 G\"oteborg, Sweden} 
\address[darmstadt]{Institut f\"ur Kernphysik, 
Technische Universit\"at Darmstadt, 64289 Darmstadt, Germany}
\address[emmi]{ExtreMe Matter Institute EMMI, GSI Helmholtzzentrum f\"ur 
Schwerionenforschung, 64291 Darmstadt, Germany}
\address[utk]{Department of Physics and Astronomy, University of
  Tennessee, Knoxville, TN 37996, USA∑}
\address[ornl]{Physics Division, Oak Ridge National Laboratory, Oak
  Ridge, TN , USA}
\address[anl]{Physics Division, Argonne National Laboratory, Argonne,
  Illinois 60439, USA}

\begin{abstract}
  We analyze the effects of finite-range corrections in halo effective
  field theory for S-wave proton halo nuclei. We calculate the charge
  radius to next-to-leading order 
  and the astrophysical S-factor for low-energy proton capture
  to fifth order in the low-energy expansion. As an application,
  we confront our results with experimental data for the S-factor
  for proton capture on Oxygen-16 into the excited $1/2^+$ state of Fluorine-17. Our
  low-energy theory is characterized by a systematic low-energy
  expansion, which can be used to quantify an energy-dependent model error to be utilized in data fitting.  Finally, we show that the
  existence of proton halos is suppressed by the need for two fine
  tunings in the underlying theory.
\end{abstract}


\end{frontmatter}


\section{Introduction
  \label{sec:intro}} 
The quantitative description of both nuclear structure and reactions
on the same footing is a major challenge of contemporary nuclear theory.
With new experimental facilities such as FRIB and FAIR at the horizon,
the task to find improved approaches for nuclear reactions
has become very urgent. \textit{Ab initio} approaches to calculate
nuclear scattering observables are limited by the computational
complexity of the nuclear many-body problem. Scattering models perform
well but use a number of uncontrolled approximations that make the
errors of such calculations difficult to quantify.

Faced with these problems, it is important to note that there are a
number of systems in the chart of nuclei for which 
the effective number of degrees-of-freedom is significantly
smaller than the number of nucleons. This phenomenon is known as clustering,
with alpha clustering, e.g. in the Hoyle state of $^{12}$C, being the 
most prominent example.
Clustering becomes even more extreme for so-called halo
nuclei which consist of a tightly-bound core nucleus and a few weakly-bound
valence nucleons \cite{Zhukov-93,Riisager:1994zz,Jensen-04,Hansen:1995pu}.
This reduction in the number of degrees of freedom is the
signature of a separation of scales in the system. 
In the case of a one-nucleon halo
nucleus, the scale separation is manifest in the small ratio of the
one-nucleon separation energy and the binding and excitation energies
of the core. For typical momenta on the order of the 
one-nucleon separation energy,
it allows for a systematic low-energy expansion in the ratio
of these two scales. 
This expansion can then be employed to calculate nuclear
observables in a model-independent and systematically improvable
manner. This approach is called halo effective field theory (Halo EFT)
\cite{Bertulani:2002sz} when a
field-theoretical approach is used for the construction of the
interaction and the calculation of observables. Halo EFT employs the
minimal number of degrees-of-freedom (core and valence nucleons) and
parameterizes the interaction in terms of a few measurable
parameters. In addition, so-called core polarization effects become important if
the core has low-lying excited states. These can be taken into
account by including excited states of the core as 
explicit degrees of
freedom in the effective theory.

Neutron halo nuclei occur rather frequently in the chart of nuclei
along the neutron dripline and have been studied in Halo EFT
\cite{Bertulani:2002sz,Bedaque:2003wa,Hammer:2011ye,Acharya:2013nia,Rupak:2012cr,Hagen:2013jqa}. Proton
halo systems exist too, but are less common due to the delicate
interplay between attraction from the strong interaction and the
repulsion from the Coulomb interaction. The effects of the Coulomb
interaction were first included into an EFT with contact interactions
by Kong and Ravndal \cite{Kong:1999sf}. Proton halos were considered
recently in
Refs. \cite{Ryberg:2013iga,Zhang:2014zsa,Ryberg:2014exa}. The Coulomb
interaction introduces a new scale into the problem that can be
understood as a result of the presence of a Coulomb barrier. This new scale is refered to as the Coulomb momentum and it is given by the inverse of the Bohr radius of the system. The
introduction of a Coulomb momentum can complicate the power counting
since it interferes with the separation of scales. Higa {\it et
al.}~\cite{Higa:2008dn}, e.g., treated the Coulomb momentum as a high-momentum
scale in their study of $\alpha-\alpha$ scattering. However, this
treatment is not always appropriate. The correct scaling of the
Coulomb momentum will always depend on the system to be
considered.

In this paper, we will extend the calculation performed in
Ref.~\cite{Ryberg:2013iga} by including higher-order
effects due to the finite range of the interaction between core and
proton. We also consider higher-order electromagnetic interactions. Specifically, we will consider the charge radius of S-wave
halo nuclei and radiative proton capture into a halo state. Our
analysis of finite-range effects also addresses the question of
why there are more neutron halos than proton halos in nature.

This manuscript is organized as follows. In Sec.~\ref{sec:theory}, we
introduce Halo EFT for S-wave systems and explain how the Coulomb
interaction is included into calculations.  In
Sections~\ref{sec:charge-form-factor} and \ref{sec:radiative-capture},
we apply Halo EFT to calculate the charge radius of proton halo nuclei
and radiative proton capture, respectively. We address the
aforementioned issue of fine tuning in proton halo nuclei in
Sec.~\ref{sec:finetuning}. Results for the excited $1/2^+$ state 
of Fluorine-17 are presented
in Sec.~\ref{sec:results-fluorine-17}. In particular, we extract the
threshold S-factor for radiative proton capture on ${}^{16}$O
into  $\nuc{17}{F}^*$
and the corresponding asymptotic normalization coefficient. To this aim we employ an order-by-order fit to experimental
radiative capture data at finite energies and we demonstrate how to quantify theoretical uncertainties within Halo EFT.  Finally, we summarize our
findings in Sec.~\ref{sec:conclusions}.

\section{Theory
\label{sec:theory}}
In Halo nuclei, the core and the valence nucleons are the effective
degrees of freedom. The Halo EFT Lagrangian can therefore be
constructed using only a core and a nucleon field. For proton halos
this was first done in Ref.~\cite{Ryberg:2013iga}.  In order to
simplify the inclusion of finite range effects and future extensions
to two-proton halos, we use an equivalent
approach that introduces an auxiliary {\it halo} field $d$ with the
quantum numbers of the halo nucleus~\cite{Bertulani:2002sz}, leading
to the Lagrangian
\begin{eqnarray}
\mathcal{L}&=&\sum_{k=0,1}\psi_k^\dagger\left[i\mathrm{D}_0+\frac{\mathrm{\mathbf{D}}^2}{2m_k}\right]\psi_k
+d^\dagger\left[\Delta+\nu\left(i\mathrm{D}_0+\frac{\mathrm{\mathbf{D}}^2}{2M_\mathrm{tot}}\right)\right]d
\nonumber
\\
&&\qquad-g\left[\psi_1^\dagger\psi_0^\dagger d+\mathrm{h.c.}\right]+\ldots
\label{eq:Lag}
\end{eqnarray}
The Lagrangian including nucleon and core fields only can be obtained
by integrating out the halo field using the classical 
equations of motion.
In Eq.~(\ref{eq:Lag}), $\psi_0$ denotes the proton field with mass $m_0$ and
$\psi_1$ the core field with mass $m_1$, while $g$ and $\Delta$ are the
low-energy constants of the theory and $M_\mathrm{tot}=m_1+m_2$. 
The parameter $\nu=\pm1$ allows
for the effective range to be both negative and positive, since we
define the coupling $g$ as purely real.\footnote{Note that the kinetic
term of the halo field $d$ has the wrong sign for $\nu=-1$
and becomes a ghost field. This introduces no pathologies here since
the $d$ field never appears in loops.}
The ellipsis denote operators with
more fields and/or additional derivatives. The covariant derivative is
$\mrm{D}_\mu=\partial_\mu+i e \hat{\mrm{Q}} A_\mu$, where
$\hat{\mrm{Q}}$ is the charge operator and $e>0$ is the elementary electric charge.  
Low-lying excitations of the
core can be included explicitly in the calculation by introducing
additional core fields (See, e.g., 
Refs.~\cite{Hammer:2011ye,Zhang:2014zsa,Ryberg:2014exa} for more details).
In the calculations below, we do not need to keep track of the proton
spin since the core-proton interaction does not change the spin of the
proton.

It is important to assign proper scaling dimensions to the fields and
operators of the Lagrangian (\ref{eq:Lag}) such that we can organize
the theory in a power counting. The proton and core fields have the
scaling dimension $\psi_k\sim 3/2$ and the scaling dimension of the
halo field is $d\sim 2$ (see for example \cite{Hammer:2011ye} for a
discussion of the scaling dimension of such auxiliary fields). The
covariant derivatives scale as $\mathrm{D}_0\sim 2$ and
$\mathrm{D}_i\sim 1$. Note that we do not explicitly 
count powers of mass factors
since mass is not equivalent to energy in non-relativistic physics.

\begin{figure}[t]
\centerline{\includegraphics*[scale=0.6,angle=0,clip=true]{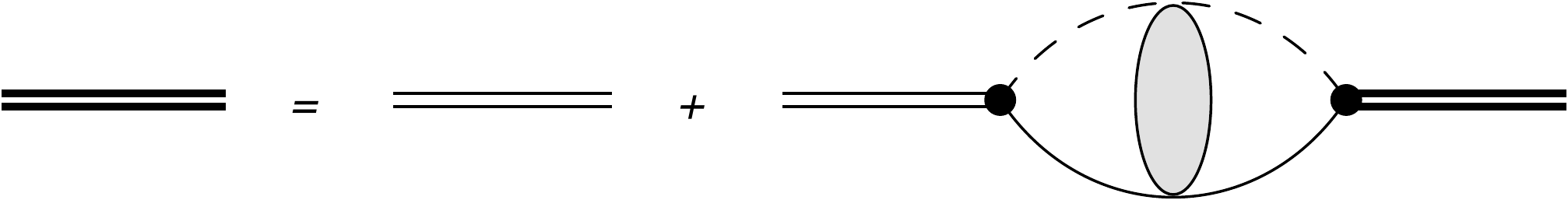}}
\caption{The full halo propagator defined iteratively. The full halo
  propagator is denoted by the thick double line. The thin double line
  denotes the bare halo propagator, the dashed single line denotes the
  core field, the solid single line denotes the proton field and the
  shaded blob the Coulomb four-point function $\chi$ defined in 
  Fig.~\ref{fig:fourpointchi}.}
\label{fig:FullDibaryon}
\end{figure}

The Feynman rules for the interactions of the Lagrangian
(\ref{eq:Lag}) are as follows: 
\begin{itemize}
\item[(i)] The vertex factor for the strong contact interaction is
$-ig$. 
\item[(ii)] The vertex factors for the $A_0$ interaction with the proton,
core, and halo field are $-ie$, $-ieZ_\mrm{c}$ and
$-i\nu e (Z_\mrm{c}+1)$, respectively, where $\Zc$ is the proton number of the core. 
\item[(iii)] The vertex factors for the interaction of the vector photon, 
$A_i$, with a proton, core, and halo field 
carrying incoming momentum  $\pp$ are
$-ie\pp/m_0$, $-ieZ_\mrm{c}\pp/m_1$ and
$-i\nu e(Z_\mrm{c}+1)\pp/M_\mrm{tot}$, respectively.
\end{itemize}

From the Lagrangian~(\ref{eq:Lag}), the one-particle propagator
can be deduced to be
\begin{equation}
  \label{eq:protonprop}
  iS_{k}(p_0,\pp)=i\left[p_0-\frac{\pp^2}{2m_k}+i\varepsilon\right]^{-1}~.
\end{equation}
For convenience, we will also define the proton-core two-particle
propagator
\begin{equation}
  iS_{\rm tot}(p_0,\pp)=i\left[ p_0-\frac{\pp^2}{2 m_{\rm R}}+i\varepsilon\right]^{-1}~,
\label{eq:Stot}
\end{equation}
where $m_{\rm R}$ denotes the reduced mass of the proton-core system.
In this work, we will only need the halo propagator
for a halo field at rest:
\begin{equation}
iD^{(0)}(E,{\mathbf{0}})\equiv
iD^{(0)}(E)=\frac{i}{\Delta+\nu\left(E+i\varepsilon\right)}~.
\end{equation}
The corresponding propagator for finite momentum $\pp$ can always be
obtained by replacing $E\to E-\pp^2/(2M_\mrm{tot})$.  The power
counting for systems interacting through a large scattering length
requires that the S-wave interaction is summed up to all orders
\cite{vanKolck:1998bw,Kaplan:1998tg,Kaplan:1998we}. The resulting full
halo propagator is thus given by the integral equation shown in
Fig.~\ref{fig:FullDibaryon}. For a halo field at rest, we obtain
\begin{equation}
iD(E)=\frac{i}{\Delta+\nu\left(E+i\varepsilon\right)+\Sigma(E)}~.
\label{eq:FullDibProp}
\end{equation}
The irreducible self-energy, $\Sigma$, includes strong and
Coulomb interactions and will be discussed below.

\begin{figure}[t]
\centerline{\includegraphics*[scale=0.6,angle=0,clip=true]{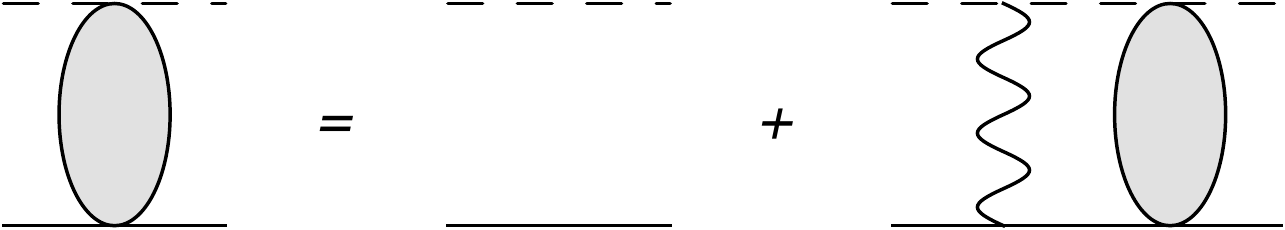}}
\caption{The four-point function $\chi$ defined iteratively. The
  wiggly line denotes a Coulomb photon exchange. 
  External propagators are amputated.
  Otherwise, the notation is as in Fig.~\ref{fig:FullDibaryon}.}
\label{fig:fourpointchi}
\end{figure}

We include the Coulomb interaction between the core and the valence
proton through the full Coulomb Green's function,
\begin{equation}
\langle \kk | G_\mrm{C}(E)|\pp\rangle = -S_{\rm
  tot}(E,\kk)\chi(\kk,\pp;E) S_{\rm tot}(E,\pp)~,
\label{eqCGFGamma}
\end{equation}
where $\pp$ and $\kk$ are the relative 
incoming and outgoing momenta and $E$ is the energy.
Here, $\chi$ is the momentum-space 
Coulomb four-point function in the center-of-mass
frame of the proton and the core, defined recursively in
Fig.~\ref{fig:fourpointchi}. To distinguish coordinate-space from
momentum-space states we will denote the former with round brackets,
i.e. $|\rr)$. The Coulomb Green's function can be expressed via its
spectral representation in coordinate space
\begin{equation}
(\rr|G_\mrm{C}(E)|\rr')=\int\frac{\d^3p}{(2\pi)^3}
\frac{\psi_\pp(\rr)\psi^*_\pp(\rr')}{E-\pp^2/(2m_\mrm{R})+i\varepsilon}~,
\label{eq:CGFSpectral}
\end{equation}
where we define the Coulomb wave function through its partial
wave expansion
\begin{equation}
\psi_\pp(\rr)=\sum_{l=0}^{\infty}(2l+1)i^l\exp{(i\sigma_l)\frac{F_l(\eta,\rho)}{\rho}P_l(\hat{\pp}\cdot\hat{\rr})}~.
\end{equation}
Here we have introduced $\rho=pr$ and the Sommerfeld parameter
$\eta= k_\mrm{C}/p$, with the Coulomb momentum
$k_\mrm{C}=Z_\mrm{c}\alpha m_\mrm{R}$, where $\alpha$ is the fine structure
constant. We have also introduced the pure Coulomb phase shift
$\sigma_l=\arg{\Gamma(l+1+i\eta)}$. For the Coulomb functions $F_l$
and $G_l$, we use the conventions of Ref.~\cite{Koenig:2012bv}.  The
regular Coulomb function $F_l$ can be expressed in terms of the
Whittaker M-function according to
\begin{equation}
F_l(\eta,\rho)=A_l(\eta)M_{i\eta,l+1/2}(2i\rho)~,
\label{eq:CoulWavefunctionF}
\end{equation}
with the $A_l$ defined as
\begin{equation}
A_l(\eta)=\frac{|\Gamma{(l+1+i\eta)}|\exp{\left[-\pi\eta/2-i(l+1)\pi/2\right]}}{2(2l+1)!}~.
\end{equation}
We will also need the irregular Coulomb wave function, $G_l$, which is given by
 \begin{equation}
G_l(\eta,\rho)=iF_l(\eta,\rho)+B_l(\eta)W_{i\eta,l+1/2}(2i\rho)~,
\end{equation}
where $W$ is the Whittaker W-function and the coefficient $B_l$ is defined as
\begin{equation}
B_l(\eta)=\frac{\exp{(\pi\eta/2+il\pi/2)}}{\arg{\Gamma{(l+1+i\eta)}}}~.
\end{equation}
It is important to note that since we consider both free and bound
states the absolute value and the argument of the $\Gamma$-function are
given by
\begin{equation}
|\Gamma{(l+1+i\eta)}|=\sqrt{\Gamma{(l+1+i\eta)}\Gamma{(l+1-i\eta)}}
\end{equation}
and
\begin{equation}
\arg{\Gamma{(l+1+i\eta)}}=\sqrt{\frac{\Gamma{(l+1+i\eta)}}{\Gamma{(l+1-i\eta)}}}~.
\end{equation}

\begin{figure}[t]
\centerline{\includegraphics*[scale=0.5,angle=0,clip=true]{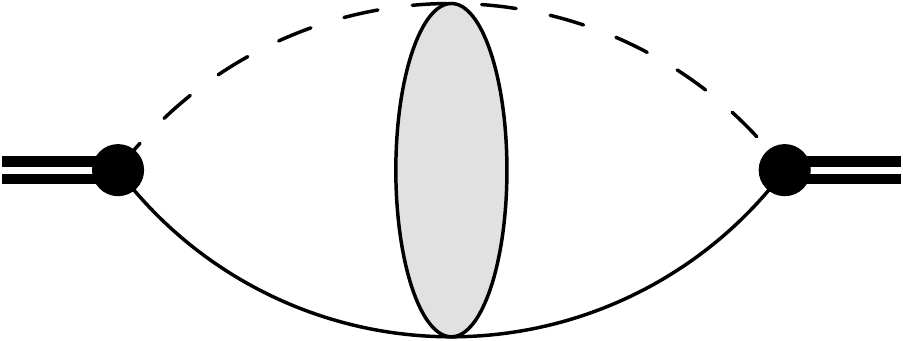}}
\caption{Feynman diagram for the irreducible self energy. The injected
  four-momentum is $(E,{\mathbf{0}})$. External legs are amputated.
  Otherwise, the notation is as in 
  Fig.~\ref{fig:FullDibaryon}.}
\label{fig:IrreducibleSelfEnergy}
\end{figure}

To obtain the fully-dressed two-particle propagator $D$, which includes
strong and Coulomb interactions, we calculate the irreducible
self-energy shown in Fig.~\ref{fig:IrreducibleSelfEnergy}. 
Using Eq.~(\ref{eqCGFGamma}), it can be expressed
by the momentum-space integral
\begin{equation}
i\Sigma(E)=-ig^2\int\frac{\d^3k_1\d^3k_2}{(2\pi)^6}~\langle \kk_2 | G_\mrm{C}(E)|\kk_1\rangle~,
\end{equation}
which can be written in coordinate space using Fourier
transformations:
\begin{align}
i\Sigma(E)=&-ig^2(0|G_\mrm{C}(E)|0)\nonumber\\
=&-ig^2\int\frac{\d^3p}{(2\pi)^3}
\frac{\psi_\pp(0)\psi^*_\pp(0)}{E-\pp^2/(2m_\mrm{R})+i\varepsilon}~.
\end{align}
We evaluate this integral using dimensional regularization in the
power divergence subtraction (PDS) scheme \cite{Kaplan:1998tg} and the
result is \cite{Kong:1999sf}
\begin{equation}
\Sigma(E)=g^2\frac{k_\mathrm{C}m_\mathrm{R}}{\pi}H(\eta)+\Sigma^\mathrm{div}~,
\end{equation}
with
\begin{equation}
H(\eta)=\psi(i\eta)+\frac{1}{2i\eta}-\log{(i\eta)}~,
\end{equation}
where $\psi(z)=\Gamma'(z)/\Gamma(z)$ is the logarithmic derivative
of the Gamma function.
The divergent part is given by
\begin{equation}
\Sigma^\mathrm{div}=-\frac{g^2k_\mathrm{C}m_\mathrm{R}}{\pi}
\Big[\frac{1}{3-d}+\log{\left(\frac{\sqrt{\pi}\mu}{2k_\mathrm{C}}\right)}
+1-\frac{3C_\mathrm{E}}{2}\Big]+\frac{g^2m_\mathrm{R}\mu}{2\pi}~,
\end{equation}
with the renormalization scale $\mu$. Note that $\Sigma^\mrm{div}$
is independent of energy and therefore will vanish when we take the
energy derivative of the irreducible self-energy to arrive at the LSZ
residue below.

\subsection{Renormalization}

\begin{figure}[t]
\centerline{\includegraphics*[scale=0.5,angle=0,clip=true]{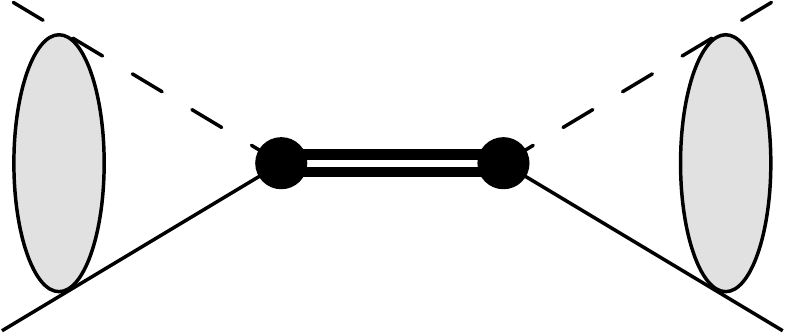}}
\caption{Scattering amplitude for elastic proton-core scattering. 
The notation is as in Fig.~\ref{fig:FullDibaryon}.}
\label{fig:ElasticScattering}
\end{figure}

Expressions for EFT low-energy constants such as $g$ and $\Delta$
defined above are frequently obtained by matching them to elastic
scattering observables. The amplitude for elastic proton-core
scattering is obtained from the diagram shown in
Fig.~\ref{fig:ElasticScattering}. It evaluates to the S-wave t-matrix
\begin{equation}
  iT_0(E)=ig^2\exp{(2i\sigma_0)}C_\eta^2D(E)~,
\label{eq:tMatrixA}
\end{equation}
with the Gamow-Sommerfeld factor $C_\eta^2=\Gamma{(1+i\eta)}\Gamma{(1-i\eta)}$. 

In the absence of the Coulomb interaction, the t-matrix is usually
expressed in terms of effective range parameters. However, it is not
possible to separate the strong interaction from the Coulomb
interaction in a model-independent way. Therefore one uses the
so-called Coulomb-modified effective range expansion (ERE) \cite{PhysRev.76.38} to relate the phase shifts to redefined effective range
parameters. The t-matrix is then written
\begin{equation}
   T_0(E)=-\frac{2\pi}{\mR}\frac{C_\eta^2\exp{(2i\sigma_0)}}{kC_\eta^2(\cot{\delta_0}-i)}~,
\label{eq:tMatrixB}
\end{equation}
where the total phase shift is given by $\sigma_0+\delta_0$.  The
S-wave Coulomb-modified ERE is
\begin{eqnarray}
kC_\eta^2(\cot{\delta_0}-i)+2\kC H(\eta)=-\frac{1}{a_0}+\frac{1}{2}r_0k^2+\ldots~,
\label{eq:CoulModERE}
\end{eqnarray}
where $a_0$ and $r_0$ are the Coulomb-modified scattering length and
effective range, respectively. One should note that the imaginary part
of $2k_\mathrm{C}H(\eta)$ exactly cancels $-ikC_\eta^2$.

Comparing Eqs.~(\ref{eq:tMatrixA}) and (\ref{eq:tMatrixB}), order
by order in the momentum $k$, we can express the scattering length and
the effective range in terms of the low-energy coupling constants $g$ and $\Delta$
\begin{align}
\label{eq:a-renorm}
\frac{1}{a_0}=&\frac{2\pi}{g^2\mR}\left(\Delta+\Sigma^\mrm{div}\right)~,\\
r_0=&-\frac{2\pi\nu}{g^2\mR^2}~.
\label{eq:r-renorm}
\end{align}
Equation (\ref{eq:a-renorm}) shows how the low-energy constant
$\Delta$ absorbs the divergent part of the irreducible self-energy
$\Sigma^\mathrm{div}$. Furthermore, at LO we effectively take $\nu$ to
zero. The two parameters $g$ and $\Delta$ are then not
independent.

The residue at the bound state, $E=-B$, of the full halo
propagator defines the wavefunction renormalization, or the LSZ
residue, and is therefore required for the calculation of bound-state
observables. It is given by
\begin{align}
\mathcal{Z}=&\left.\left[\frac{\d\left(D^{-1}\right)}{\d E}\right]^{-1}\right|_{E=-B}
=\frac{1}{\nu+\Sigma'(-B)}~.
\label{eq:LSZRes}
\end{align}
In terms of the effective range, we write this as
\begin{equation}
\mathcal{Z}=\frac{6\pi\kC}{g^2\mR^2}\frac{1}{\tilde{H}(\gamma,\kC)-3\kC r_0}~,
\label{eq:LSZRes1}
\end{equation}
using the matching condition in Eq.~(\ref{eq:r-renorm}). In writing
Eq.~(\ref{eq:LSZRes1}), we have defined the function
\begin{equation}
\tilde{H}(\gamma,\kC)=\left.\frac{6\kC^2}{\mR}\frac{\d}{\d E}H(\eta)\right|_{E=-B}~,
\end{equation}
with the binding momentum $\gamma=\sqrt{2\mR B}$.
The expression Eq.~(\ref{eq:LSZRes1}) is valid at next-to-leading order
(NLO), that is it includes the effective-range correction. At leading
order (LO) the wavefunction renormalization is given by the simpler
expression
\begin{align}
\mathcal{Z}_\mrm{LO}=&\frac{1}{\Sigma'(-B)}
\label{eq:LSZSigmaLO}\\
=&\frac{6\pi\kC}{g^2\mR^2}\frac{1}{\tilde{H}(\gamma,\kC)}
\label{eq:ZLO}
\end{align}
that is Eq.~(\ref{eq:LSZRes1}) with $r_0$ set to zero.  Note that the
factor of $g^2$ in the denominator of the wavefunction renormalization
will always cancel with a corresponding factor in an unrenormalized
matrix element. 

If the ratio between the binding momentum and Coulomb momentum
$\gamma/k_\mathrm{C}$ is small we can expand the function $\tilde{H}(\gamma,\kC)$ according to
\begin{equation}
\tilde{H}(\gamma,\kC)=1-\frac{\gamma^2}{5\kC^2}+\frac{\gamma^4}{7\kC^4}+\dots~.
\end{equation}
Thus, for systems where the separation $\gamma\ll\kC$ is fulfilled, we
can use $\tilde{H}(\gamma,\kC)\to1$ in all expressions.
(See Ref.~\cite{Higa:2008dn} for a similar expansion of $H(\eta)$.)

 In this paper, we will fix the LSZ residue $\mathcal{Z}$ to a
 calculated or measured asymptotic normalization coefficient (ANC),
 $A$, or to experimental radiative capture data. The ANC is defined as the coefficient in the asymptotic bound
 state wavefunction
\begin{equation}
w_l(r)=AW_{-i\eta,l+1/2}(2\gamma r)~,
\end{equation}
where $W$ is the Whittaker-W function. The LSZ residue is related to the ANC according to
\begin{equation}
\mathcal{Z}=\frac{\pi}{g^2\mR^2\left[\Gamma(1+\kC/\gamma)\right]^2}A^2~.
\label{eq:Z-ANC}
\end{equation}
At LO, the wavefunction renormalization is determined solely by
$\gamma$ and $\kC$, as can be seen in Eq.~(\ref{eq:ZLO}), and as such
this could in principle be used to predict the LO ANC as
\begin{equation}
A_\mrm{LO}=\sqrt{\frac{6\kC}{\tilde{H}(\gamma,\kC)}}\Gamma(1+\kC/\gamma)~.
\label{eq:LOANC}
\end{equation}

At NLO we are left with one undetermined parameter, $r_0$, in the LSZ
residue~(\ref{eq:LSZRes1}). Thus, at orders
beyond LO we can use an ANC as input to fix $\mathcal{Z}$ and to
predict the effective-range parameters. We will therefore define our
NLO wavefunction renormalization in terms of the matching to the ANC,
Eq.~(\ref{eq:Z-ANC}). The ratio to the LO residue is then obtained as
\begin{equation}
\frac{\mathcal{Z}}{\mathcal{Z}_\mrm{LO}}=
\frac{\tilde{H}(\gamma,\kC)A^2}{6\kC\left[\Gamma(1+\kC/\gamma)\right]^2}~.
\label{eq:ZNLOoverZLO}
\end{equation}
Note that the matching (\ref{eq:Z-ANC}) is valid at any order in the
power counting and as such there is no EFT error due to the
non-inclusion of higher-order contact interactions.

For a given one-proton separation energy, the ANC determines the
Coulomb-modified effective range
\begin{align}
  \label{eq:anc-range}
  r_0=&\frac{1}{3\kC}\left[\tilde{H}(\gamma,\kC)-\frac{6 \kC \Gamma(1+\kC/\gamma)^2}{A^2}\right]
\end{align}
combining Eqs.~(\ref{eq:LSZRes1}) and (\ref{eq:Z-ANC}). The Coulomb-modified scattering length is then obtained from the pole position of the t-matrix (\ref{eq:tMatrixB}), that is 
\begin{equation}
a_0=-\frac{2}{4\kC H(-i\kC/\gamma)+\gamma^2r_0}~.
\end{equation}
These S-wave effective-range parameter predictions are accurate up to corrections of the shape parameter in the ERE.

\section{Charge form factor}
\label{sec:charge-form-factor}
Information on the electromagnetic structure of an object can be
obtained through elastic electron scattering. The elastic scattering
amplitude can then be expressed in terms of the electric and magnetic
form factors. Here we will focus on the electric charge form factor of an
S-wave halo nucleus. The charge form factor, $F_\mrm{C}$, and charge
radius, $r_\mrm{C}$, are defined by
\begin{align}
F_\mathrm{C}(Q)=&\frac{1}{e(\Zc+1)}\langle\pp'|J_\mathrm{EM}^0|\pp\rangle\\
=&1-\frac{r_\mathrm{C}^2}{6}Q^2+\ldots
\end{align}
where $\pp$ ($\pp'$) is the incoming (outgoing) momentum of the halo
field, $\QQ=\pp'-\pp$ is the momentum transfer, and $J_\mrm{EM}^\mu$ is
the electromagnetic current. We evaluate this observable in the Breit
frame, where no energy is transferred such that the photon
four-momentum is $(0,\QQ)$.

In this paper, we calculate the charge form factor to NLO, by
evaluating the relevant diagrams to this order. At LO there are two
loop diagrams, $\Gamma_\mrm{LO}(Q)$, and at NLO a constant tree-level
diagram, $\Gamma_\mrm{NLO}$, enters. We derive and evaluate these
diagrams below. The charge form factor is then given by the sum of
diagrams
\begin{equation}
F_\mrm{C}(Q)=\frac{1}{e(\Zc+1)}\mathcal{Z}\left[\Gamma_\mrm{LO}(Q)
+\Gamma_\mrm{NLO}(Q)+\dots\right]~.
\label{eq:ChargeFF3}
\end{equation}

\subsection{Leading order}

\begin{figure}[t]
\centerline{\includegraphics*[scale=0.5,angle=0,clip=true]{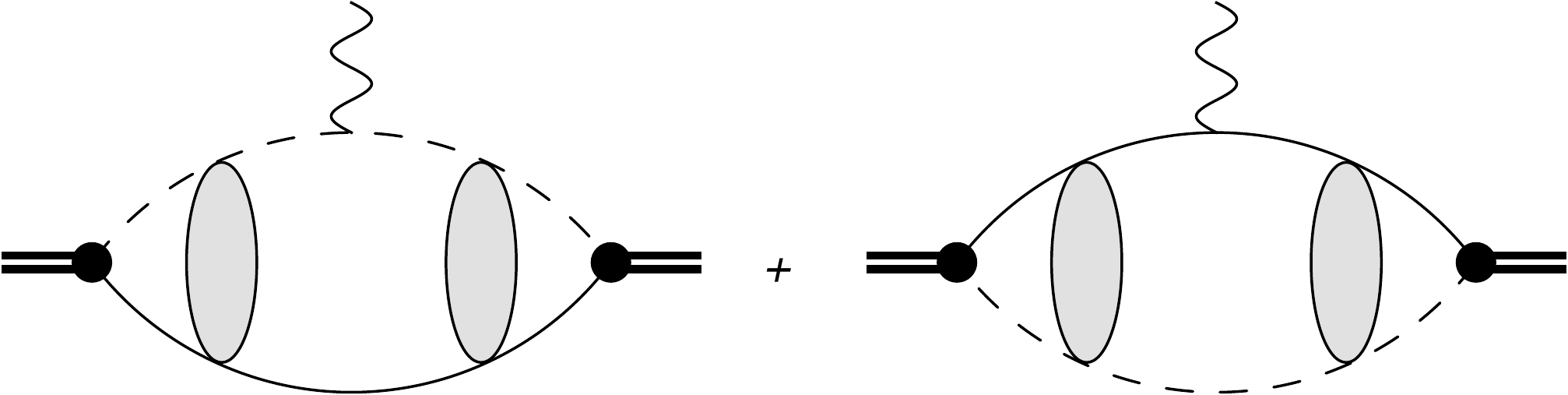}}
\caption{The diagrams for the charge form factor at LO. The notation is as in
Fig.~\ref{fig:FullDibaryon}.}
\label{fig:GammaLO}
\end{figure}

At leading order, we have to consider the diagrams shown in
Fig.~\ref{fig:GammaLO}. There, the photon couples to the
single-particle lines only, that is through an operator 
$\psi^\dagger_kA_0\psi_k$ of dimension $5$. 
We choose incoming and
outgoing total four-momenta as $(E,-\QQ/2)$ and $(E,\QQ/2)$,
respectively. The resulting amplitude in momentum space is given by
\begin{align}
i\Gamma_\mrm{LO}(Q)=&ig^2e\Zc\int\frac{\d^4k_1\d^4k_2\d^4k_3}{(2\pi)^{12}}~iS_0(k_{30},\kk_3)\nonumber\\
&\times iS_1(E-k_{30},-\kk_3+\QQ/2)~i\chi(\kk_3-f\QQ/2,\kk_2-f\QQ/2,-B)\nonumber\\
&\times iS_0(k_{20},\kk_2)~iS_1(E-k_{20},-\kk_2+\QQ/2)\nonumber\\
&\times iS_1(E-k_{20},-\kk_2-\QQ/2)~i\chi(\kk_2+f\QQ/2,\kk_1+f\QQ/2,-B)\nonumber\\
&\times iS_0(k_{10},\kk_1)~iS_1(E-k_{10},-\kk_1-\QQ/2)\nonumber\\
&+\left[(f\to1-f),~(\Zc\to 1)~,~(S_0\leftrightarrow S_1)\right]\,,
\label{eq:GammaLOmom1}
\end{align}
where $f=m_0/\Mtot$ is a mass ratio introduced for convenience. The
last row makes sure that the photon also couples to the proton. Note
that the total energy $E$ is given by $E=-B+Q^2/(8\Mtot)$, using
energy conservation.

We evaluate the energy integrals in Eq.~(\ref{eq:GammaLOmom1}) using
the residue theorem, noting that the poles are located at
$k_{n0}=-i\varepsilon+\kk_{n}^2/(2m_0)$ for $n=1,2,3$. The result can
be written using two-body propagators, defined in Eq.~(\ref{eq:Stot}),
as
\begin{align}
i\Gamma_\mrm{LO}(Q)=&-ig^2e\Zc\int\frac{\d^3k_1\d^3k_2\d^3k_3}{(2\pi)^{9}}~\Stot(-B,\kk_3)\nonumber\\
&\times\chi(\kk_3,\kk_2-f\QQ/2,-B)~\Stot(-B,\kk_2-f\QQ/2)\nonumber\\
&\times\Stot(-B,\kk_2+f\QQ/2)~\chi(\kk_2+f\QQ/2,\kk_1,-B)\nonumber\\
&\times\Stot(-B,\kk_1)\nonumber\\
&+\left[(f\to1-f),~(\Zc\to 1)\right]\,,
\end{align}
and this expression can be simplified using the Coulomb Green's
function in Eq.~(\ref{eqCGFGamma}), to replace the two-body
propagators $\Stot$ and the four-point function $\chi$. This leads to
\begin{align}
i\Gamma_\mrm{LO}(Q)=&-ig^2e\Zc\int\frac{\d^3k_1\d^3k_2\d^3k_3}{(2\pi)^{9}}~\langle\kk_3|G_\mrm{C}(-B)|\kk_2-f\QQ/2\rangle\nonumber\\
&\times\langle\kk_2+f\QQ/2|G_\mrm{C}(-B)|\kk_1\rangle\nonumber\\
&+\left[(f\to1-f),~(\Zc\to 1)\right]~.
\label{eq:GammaLOmom2}
\end{align}

By performing a Fourier transform on each of the momentum-space bras
and kets, we arrive at the coordinate-space integral
\begin{align}
i\Gamma_\mrm{LO}(Q)=&-ig^2e\Zc\int\frac{\d^3r}{(2\pi)^3}\nonumber\\
&\times(0|G_\mrm{C}(-B)|\rr)~\exp{(if\QQ\cdot\rr)}~(\rr|G_\mrm{C}(-B)|0)\nonumber\\
&+\left[(f\to1-f),~(\Zc\to 1)\right]~.
\label{eq:GammaLOcoord1}
\end{align}
This integral is much more convenient to use than the rather involved
mo\-men\-tum-space integral in Eq.~\eqref{eq:GammaLOmom2}. In the integral
(\ref{eq:GammaLOcoord1}) the diagram in Fig.~\ref{fig:GammaLO} is also
visualized better. It consists of two Coulomb Green's functions,
that propagate the fields from separation zero to $\rr$ and back from 
separation $\rr$ to zero, respectively, and the current operator in between
the propagators. Since the Coulomb Green's functions have one end at
zero separation, only the S-wave part of these will
contribute. Therefore we expand the Coulomb Green's function in
partial waves
\begin{equation}
(\rr'|G_\mrm{C}(E)|\rr)=\sum_{l=0}^\infty(2l+1)G^{(l)}_\mrm{C}(E;r',r)P_l(\hat{\rr}'\cdot\hat{\rr})~,
\end{equation}
with the bound-state partial-wave projected Coulomb Green's function
\begin{equation}
G^{(l)}_\mrm{C}(-B;r',r)=-i\frac{\mR\gamma}{2\pi}\frac{F_l(\eta,\rho')
\left[iF_l(\eta,\rho)+G_l(\eta,\rho)\right]}{\rho'\rho}~.
\end{equation}
We can now write the Green's function as (see \ref{sec:app} for details)
\begin{align}
(0|G_\mrm{C}(-B)|\rr)=&G^{(0)}_\mrm{C}(-B;0,r)\nonumber\\
=&-\frac{\mR\Gamma(1+\kC/\gamma)}{2\pi}\frac{W_{-\kC/\gamma,1/2}(2\gamma r)}{r}~.
\end{align} 
As such, the integral (\ref{eq:GammaLOcoord1}) can be written as
\begin{align}
i\Gamma_\mrm{LO}(Q)=&-i\frac{g^2e\Zc\mR^2}{8\pi^4}
\Gamma(1+\kC/\gamma)^2\int\d r~j_0(fQr)W_{-\kC/\gamma,1/2}(2\gamma r)^2\nonumber\\
&+\left[(f\to1-f),~(\Zc\to 1)\right]~.
\label{eq:GammaLOcoord2}
\end{align}
This integral can be evaluated numerically in a straightforward
way.
The LO charge form factor can now be calculated through
\begin{equation}
F_\mrm{C}(Q)\Big|_\mrm{LO}=\frac{1}{e(\Zc+1)}\mathcal{Z}_\mrm{LO}\Gamma_\mrm{LO}(Q)~,
\label{eq:chargeformfactorLO}
\end{equation}
and the LO charge radius is given in terms of the loop-integral $\Gamma_\mrm{LO}(Q)$ and the wavefunction renormalization $\mathcal{Z}_\mrm{LO}$, according to
\begin{equation}
r_\mrm{C}^2\Big|_\mrm{LO}=-\frac{3\mathcal{Z}_\mrm{LO}}{e(\Zc+1)}\left.\frac{\d^2}{\d Q^2}\Gamma_\mrm{LO}(Q)\right|_{Q=0}~.
\end{equation}

We will now show that the charge form factor is normalized correctly
to 1 at $Q=0$. Starting from the coordinate-space integral
(\ref{eq:GammaLOcoord1}) at $Q=0$ and the spectral representation of
the Coulomb Green's function
\begin{equation}
(0|G_\mrm{C}(E)|\rr)=\int\frac{\d^3p}{(2\pi)^3}
\frac{\psi_\pp(0)\psi^*_\pp(\rr)}{E-\pp^2/(2m_\mrm{R})+i\varepsilon}~,
\label{eq:CGFSpectral2}
\end{equation}
we find that
\begin{align}
\Gamma_\mrm{LO}(0)=&-g^2e(\Zc+1)\int\d^3r\left|(0|G_\mathrm{C}(-B)|r)\right|^2\nonumber\\
=&-g^2e(\Zc+1)\int\frac{\d^3p}{(2\pi)^3}\frac{\psi_\pp(0)\psi^*_\pp(0)}{(E-\pp^2/(2m_\mathrm{R})+i\varepsilon)^2}\nonumber\\
=&e(Z_\mathrm{c}+1)\Sigma'(-B)~.
\label{eq:GammaLOlimit}
\end{align}
In the first step above, a Dirac delta was used, coming from the
integration over the orthonormal Coulomb wavefunctions
\begin{equation}
\int\d^3r~\psi^*_\pp(\rr)\psi_{\pp'}(\rr)=(2\pi)^3\delta^{(3)}(\pp-\pp')~.
\end{equation}
The correct normalization of the LO charge form factor now follows,
combining Eqs.~(\ref{eq:chargeformfactorLO}), (\ref{eq:GammaLOlimit})
and (\ref{eq:LSZSigmaLO}).

\subsection{Next-to-leading order}

\begin{figure}[t]
\centerline{\includegraphics*[scale=0.5,angle=0,clip=true]{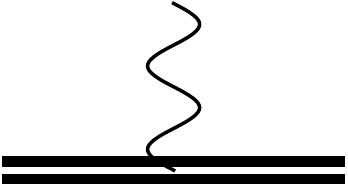}}
\caption{The diagram for the charge form factor at NLO. The notation is as in
Fig.~\ref{fig:FullDibaryon}.}
\label{fig:GammaNLO}
\end{figure}

The contributions that enter at NLO are that the full LSZ residue
(\ref{eq:LSZRes}), or (\ref{eq:Z-ANC}), is to be used and that the NLO
tree-level diagram, given in Fig.~\ref{fig:GammaNLO}, enters, through the operator $d^\dagger A_0d$ of dimension $6$. This
diagram is simply given by the Feynman rule for an $A_0$ photon
coupling to the dicluster field
\begin{equation}
i\Gamma_\mrm{NLO}=i\nu e(\Zc+1)~.
\label{eq:GammaNLO}
\end{equation}
Note that the NLO correction diagram (\ref{eq:GammaNLO}) is
independent of the momentum transfer $Q$.  These contributions come
with an effective-range correction.

The NLO charge form factor is given by the sum of the diagrams up to
NLO, according to Eq.~(\ref{eq:ChargeFF3}).
It is clear that the normalization of the charge form factor at $Q=0$
is still correct, using Eq.~(\ref{eq:GammaLOlimit}) together with the
NLO LSZ residue (\ref{eq:LSZRes}), the NLO diagram (\ref{eq:GammaNLO})
and the formula (\ref{eq:ChargeFF3}). The charge radius is
then given by the order $Q^2$ part of the LO loop-integral
(\ref{eq:GammaLOcoord2}) together with the full wavefunction
renormalization (\ref{eq:Z-ANC}). The resulting NLO charge radius
result is
\begin{equation}
r_\mrm{C}^2=-\frac{3\mathcal{Z}}{e(\Zc+1)}\frac{d^2}{dQ^2}\left.\Gamma_\mrm{LO}(Q)\right|_{Q=0}~.
\end{equation}
It is evident that the NLO charge radius squared is a factor 
\begin{equation}
\frac{\mathcal{Z}}{\mathcal{Z}_\mrm{LO}}=
\frac{\tilde{H}(\gamma,\kC)A^2}{6\kC\Gamma(1+\kC/\gamma)^2}~.
\tag{\ref{eq:ZNLOoverZLO}}
\end{equation}
larger (or smaller) than the LO result.

At higher orders there are three types of corrections. Firstly, there are local short-range operators $\psi_k^\dagger\left[\nabla^2A_0-\partial_0(\nabla\cdot\mathbf{A})\right]\psi_k$, of dimension $7$, which enter with finite-size contributions of the core and proton fields. Secondly, there is a local short-range operator $d^\dagger\left[\nabla^2A_0-\partial_0(\nabla\cdot\mathbf{A})\right]d$, of dimension $8$, that comes in with an undetermined short-range parameter. Thirdly, there are photon couplings due to the minimal substitution of derivatives in higher-order contact interactions, with the leading N$^3$LO contribution coming from the shape-parameter in the ERE.

\section{Radiative capture}
\label{sec:radiative-capture}

In this section we consider radiative capture of a proton into a halo
state. By including range corrections, we are able to compute the cross
section for this process to high orders in the Halo EFT expansion. We
use quantum numbers relevant for the capture process
$^{16}\mathrm{O}(\mathrm{p},\gamma)^{17}\mathrm{F}^*$, which require
that the incoming particle pair has relative angular momentum $l\geq1$. 
Specifically, we will consider the E1 capture through an incoming P-wave.

The differential cross section for radiative capture of
non-relativistic particles is
\begin{equation}
\frac{\d\sigma}{\d\Omega}=\frac{m_\mathrm{R}\omega}{8\pi^2p}\sum_i\left|\epsilon_i\cdot\mathcal{A}\right|^2~,
\end{equation}
where $\omega$ is the energy of the outgoing photon, $p$ is the
relative momentum of the incoming particle pair and $\epsilon_i$ are
the photon polarization vectors. The vector amplitude $\mathcal{A}$ is
for the capture process with a vector photon $A_i$ being emitted. Note
that we are working in Coulomb gauge, where the relation
\begin{equation}
\epsilon_i\cdot\QQ=0~,
\label{eq:Ward1}
\end{equation}
for a real photon with momentum $\QQ$, is fulfilled.

We present our results in terms of the astrophysical S-factor, which
is defined as
\begin{equation}
S(E)=E\exp{(2\pi\eta)}\sigma_\mathrm{tot}(E)~,
\end{equation}
with the incoming center-of-mass energy $E$ and the total cross section $\sigma_\mrm{tot}$.

From the Lagrangian (\ref{eq:Lag}), we can write down three classes of
diagrams for the radiative capture process. Two of these are
identically zero since the incoming particle pair is in a relative
S-wave. For completeness we will show this explicitly below. The
remaining diagram has contributions from partial waves $l\geq1$, but
since the P-wave dominates at low energies we neglect all other
partial waves.

\subsection{Leading order}

\begin{figure}[t]
\centerline{\includegraphics*[scale=0.5,angle=0,clip=true]{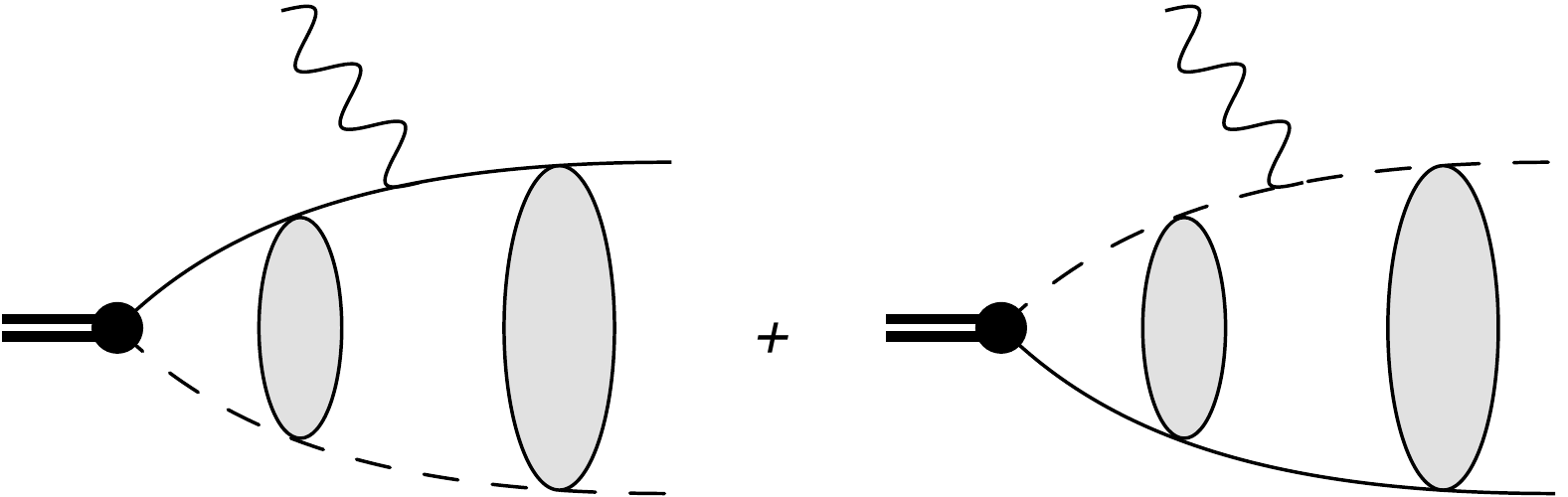}}
\caption{The diagrams for radiative capture at LO. The notation is as in
Fig.~\ref{fig:FullDibaryon}.}
\label{fig:captureLO}
\end{figure}

At leading order, we only consider diagrams where the photon couples
to one of the single-particle lines, that is through an operator
$\psi^\dagger_k\frac{ie\hat{\mrm{Q}}A_i\nabla_i}{m_k}\psi_k$ of
dimension $5$.  The two radiative capture diagrams of interest are
shown in Fig.~\ref{fig:captureLO}. We will write them using the pure
Coulomb t-matrix $T_\mrm{C}$ \cite{ChenChen:1972}. 
The four-point function $\chi$ defined in Fig.~\ref{fig:fourpointchi} is
directly proportional to the Coulomb Green's function $G_C$ and
receives contributions from the Coulomb t-matrix $T_C$ as well as from the 
free propagation of the core-proton system (cf.~Fig.~\ref{fig:fourpointchi}). 
In the center-of-mass of mass of the halo and the radiated photon,
the momentum space diagrams from  Fig.~\ref{fig:captureLO}
are given by
\begin{align}
i\mathcal{A}=&-ig\sqrt{\mathcal{Z}_\mrm{LO}}\int\frac{\d^4k_2}{(2\pi)^4}~iS_0(k_{20},\kk_2)~iS_1(E-k_{20}-\omega,-\kk_2-\QQ)\nonumber\\
&\times\bigg[i\chi(\kk_2+f\QQ,\pp+f\QQ,-B)~iS_1\left(\pp^2/(2m_1)-\omega,-\pp-\QQ\right)\nonumber\\
&\times\left(-ieZ_\mrm{c}(-\pp)/m_1\right)\nonumber\\
&+\int\frac{\d^4k_1}{(2\pi)^4}~i\chi(\kk_2+f\QQ,\kk_1+f\QQ,-B)~iS_1(E-k_{10}-\omega,-\kk_1-\QQ)\nonumber\\
&\times\left(-ieZ_\mrm{c}(-\kk_1)/m_1\right)\nonumber\\
&\times iS_1(E-k_{10},-\kk_1)~iS_0(k_{10},\kk_1)~\left(-iT_\mrm{C}(\kk_1,\pp)\right)\bigg]\nonumber\\
&-\left[(f\to1-f)~,~(\Zc\to1)~,~(m_0\leftrightarrow m_1)~,~(S_0\leftrightarrow S_1)\right]~,
\label{eq:CaptureLO0}
\end{align}
where we have separated the contributions from the  Coulomb t-matrix $T_C$
and the free propagation of the core-proton system  in the initial state.
The total energy flowing through the diagram is 
$E=-B+\omega+\omega^2/({2M_\mathrm{tot}})$ and the factor
$\sqrt{\mathcal{Z}_\mrm{LO}}$ is from the wavefunction renormalization
of the halo. Performing the
$k_{n0}$ residue integrals and introducing a convenient Dirac delta,
the integral (\ref{eq:CaptureLO0}) can be written as
\begin{align}
i\mathcal{A}=&-ig\sqrt{\mathcal{Z}_\mrm{LO}}\frac{eZ_\mrm{c}f}{\mR}\int\frac{\d^3k_1\d^3k_2}{(2\pi)^6}~S_\mrm{tot}(-B,\kk_2+f\QQ)~\chi(\kk_2+f\QQ,\kk_1+f\QQ,-B)\nonumber\\
&\times S_\mrm{tot}(-B,\kk_1+f\QQ)~\kk_1\nonumber\\
&\times\bigg[\delta^3(\pp-\kk_1)+\left[E-\kk_1^2/(2\mR)\right]^{-1}T_\mrm{C}(\kk_1,\pp)\bigg]\nonumber\\
&-\left[(f\to1-f)~,~(\Zc\to1)\right]~.
\label{eq:CaptureLO01}
\end{align}
Then, using the Lippmann-Schwinger relation \cite{ChenChen:1972}
\begin{equation}
\psi_\pp(\kk_1)=\delta^{3}(\kk_1-\pp)+\left[E-\kk_1^2/(2\mR)\right]^{-1}T_\mrm{C}(\kk_1,\pp)
\end{equation}
and replacing the four-point function $\chi$ with
Eq.~(\ref{eqCGFGamma}) the integral (\ref{eq:CaptureLO01}) can be
expressed in a simpler fashion:
\begin{align}
i\mathcal{A}=&-ig\sqrt{\mathcal{Z}_\mrm{LO}}\frac{e\Zc f}{\mR}\int\frac{\d^3k_1\d^3k_2}{(2\pi)^6}~\langle\kk_2|G_\mrm{C}(-B)|\kk_1+f\QQ\rangle~\kk_1~\psi_\pp(\kk_1)\nonumber\\
&-\left[(f\to1-f)~,~(\Zc\to1)\right]
\label{eq:CaptureLO1}
\end{align}
Performing Fourier transforms and using $\kk_1\exp{(i\kk_1\cdot\rr_2)}=-i\nabla_2\exp{(i\kk_1\cdot\rr_2)}$ we can write the integral (\ref{eq:CaptureLO1}) as 
\begin{align}
i\mathcal{A}=&-g\sqrt{\mathcal{Z}_\mrm{LO}}\frac{e\Zc f}{\mR}\int\d^3r_1\d^3r_2~G_\mrm{C}^{(0)}(-B;0,pr_1)~\psi_\pp(\rr_2)\nonumber\\
&\times\exp{(-if\QQ\cdot\rr_1)}~\nabla_2\delta^{(3)}(\rr_2-\rr_1)\nonumber\\
&-\left[(f\to1-f)~,~(\Zc\to1)\right]~.
\label{eq:CaptureLO2}
\end{align}
The resulting integral can now be expressed as 
\begin{align}
i\mathcal{A}=&g\sqrt{\mathcal{Z}_\mrm{LO}}\frac{e\Zc f}{\mR}\int\d^3r~G_\mrm{C}^{(0)}(-B;0,pr)~\exp{(-if\omega r \cos{\theta})}~\left(\nabla\psi_\pp(\rr)\right)\nonumber\\
&-\left[(f\to1-f)~,~(\Zc\to1)\right]~.
\label{eq:CaptureLO3}
\end{align}
Summing over all polarizations and  doing the angular integration,
the amplitude squared in Eq.~(\ref{eq:CaptureLO3})
evaluates to
\begin{eqnarray}
\sum_i\left|\epsilon_i\cdot\mathcal{A}\right|^2&=&\Bigg|\sqrt{\mathcal{Z}}\sin{\theta}(\cos{\phi}+\sin{\phi})\frac{4\pi ge\Zc f\exp{(i\sigma_1)}}{\mR p}\nonumber\\
&&\times\int\d r~G_\mrm{C}^{(0)}(-B;0,\rho)j_0(f\omega r)\frac{\partial}{\partial r}\left[rF_1(\kC/p,pr)\right]\nonumber\\
&&-\left[(f\to1-f)~,~(\Zc\to1)\right]\Bigg|^2~.
\end{eqnarray}
This integral can be solved numerically, using the S-wave Coulomb Green's
function given in Eq.~(\ref{eq:CGF4}) and the regular Coulomb wave
function defined in Eq.~(\ref{eq:CoulWavefunctionF}).

\begin{figure}[t]
\centerline{\includegraphics*[scale=0.5,angle=0,clip=true]{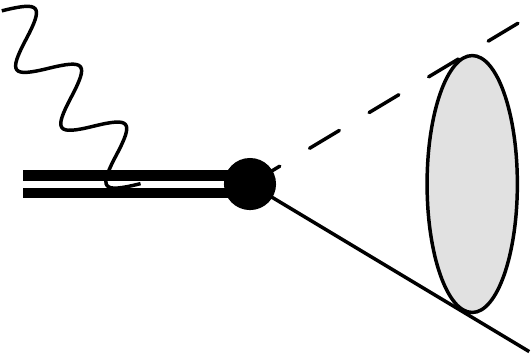}}
\caption{The first of the two vanishing classes of capture diagrams at LO. The notation is as in Fig.~\ref{fig:FullDibaryon}.}
\label{fig:CaptureNLO1}
\end{figure}

\begin{figure}[t]
\centerline{\includegraphics*[scale=0.5,angle=0,clip=true]{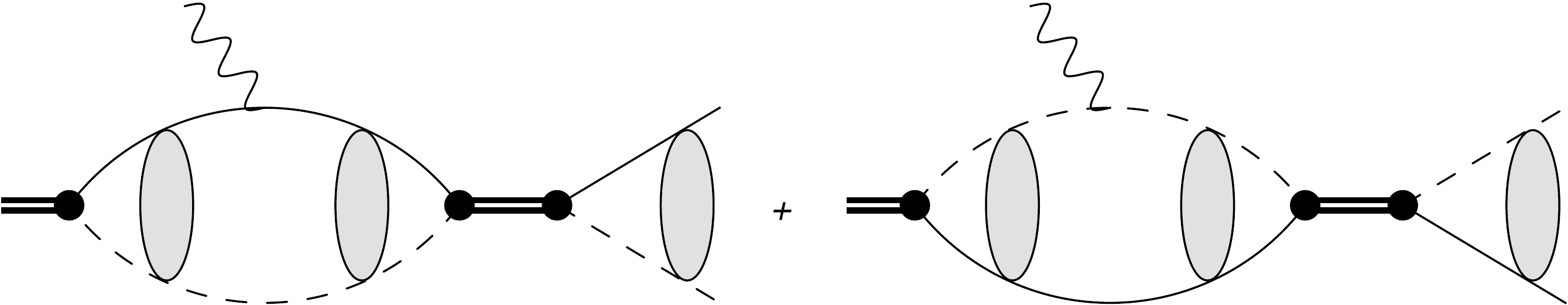}}
\caption{The second of the two vanishing classes of capture diagrams at LO. The notation is as in Fig.~\ref{fig:FullDibaryon}.}
\label{fig:CaptureNLO2}
\end{figure}

The remaining two classes of diagrams to consider at LO evaluate to
zero for our choice of reference frame and kinematics.
This can be understood from the fact that they correspond to an incoming 
S-wave. The diagram shown in
Fig.~\ref{fig:CaptureNLO1} evaluates to zero in the zero-momentum
frame, since the photon coupling is proportional to the ingoing
momentum $\pp'=0$ of the dicluster field:
\begin{align}
i\mathcal{A}^{(1)}=&-i\sqrt{\mathcal{Z}}\frac{\nu e(\Zc+1)}{M_\mrm{tot}}\pp'iD(E,\pp')(-ig)\nonumber\\
&\times\left[1+\int\frac{\d^4k}{(2\pi)^4}~iS_0(k_0,\kk)~iS_1(E-k_0,-\kk)~iT_\mrm{C}(\kk,\pp)\right]\nonumber\\
=&0
\end{align}
The momentum-space amplitude from the next diagrams, which are 
shown in Fig. \ref{fig:CaptureNLO2}, is given by
\begin{align}
i\mathcal{A}^{(2)}=&-ig\sqrt{\mathcal{Z}}\int\frac{\d^4k_1\d^4k_2\d^4k_3}{(2\pi)^{12}}~iS_0(k_{30},\kk_3)~iS_1(E-\omega-k_{30},-\kk_3-\QQ)\nonumber\\
&\times i\chi(\kk_3+f\QQ,\kk_2+f\QQ,-B)~iS_0(k_{20},\kk_2)~iS_1(E-\omega-k_{20},-\kk_2-\QQ)\nonumber\\
&\times\left(-ieZ_\mrm{c}(-\kk_2)/m_1\right)~iS_1(E-k_{20},-\kk_2)~i\chi(\kk_2,\kk_3,-B)\nonumber\\
&\times iS_0(k_{10},\kk_1)~iS_1(E-k_{10},-\kk_1)~(-ig)~iD(E,0)~(-ig)\nonumber\\
&\times\left[1+\int\frac{\d^4k}{(2\pi)^4}~iS_0(k_0,\kk)~iS_1(E-k_0,-\kk)~iT_\mrm{C}(\kk,\pp)\right]~\nonumber\\
&+\left[(f\to1-f)~,~(\Zc\to1)~,~(S_0\leftrightarrow S_1)\right]~.
\end{align}
Doing the same simplifications as before, it can be shown that this
amplitude is parallel to the photon momentum:
\begin{align}
i\mathcal{A}^{(2)}=&-ig\sqrt{\mathcal{Z}}\int\d^3r~G_\mrm{C}(-B;0,\rho)~i\frac{fe\Zc}{\mR}\exp{(-if\QQ\cdot\rr)}\nonumber\\
&\times\left[\nabla G_\mrm{C}(-B;\rho,0)\right](-ig)iD(E,0)(-ig)\psi_\pp(0)\nonumber\\
&+\left[(f\to1-f)~,~(\Zc\to1)\right]\nonumber\\
=&-ig^3\sqrt{\mathcal{Z}}D(E,0)\psi_\pp(0)\frac{e\Zc f}{\mR}\int\d^3r~G_\mrm{C}(-B;0,\rho)\nonumber\\
&\times\sum_l(2l+1)i^lj_l(f\omega r)P_l(-\hat{\QQ}\cdot\hat{\rr})\left[\hat{\rr}\frac{\partial}{\partial r}G_\mrm{C}(-B;\rho,0)\right]\nonumber\\
&+\left[(f\to1-f)~,~(\Zc\to1)\right]\nonumber\\
=&\hat{\QQ}g^3\sqrt{\mathcal{Z}}D(E,0)\psi_\pp(0)\frac{e\Zc f}{\mR}\int\d^3r~G_\mrm{C}(-B;0,\rho)j_1(f\omega r)\nonumber\\
&\times\left[\frac{\partial}{\partial r}G_\mrm{C}(-B;\rho,0)\right]\nonumber\\
&+\left[(f\to1-f)~,~(\Zc\to1)\right]
\label{eq:CaptureNLO2}
\end{align}
In the first step the exponential function is expanded in spherical
Bessel functions and in the second step all partial waves but $l=1$
integrates to zero. Thus, using Eq.~(\ref{eq:Ward1}),
\begin{equation}
\epsilon_i\cdot\mathcal{A}^{(2)}=0
\end{equation}
and the diagram does therefore not contribute.

\subsection{Next-to-leading order}
The higher-order ERE parameters appear with $\nabla + ie\hat{\mrm{Q}}
\mathbf{A}$ operators that, 
in principle, can give contributions to the radiative-capture amplitude. 
However, the diagrams with these higher-order ERE operators are diagrams 
with initial-wave scattering due to the strong force. Since we only have 
included the S-wave part of the strong interaction these initial-wave 
scattering diagrams are identically zero, which can be understood from 
the fact that the E1 capture process changes the angular momentum by one. 
If we were to include also the P-wave interaction explicitly in the field 
theory, then the effective range, the shape parameter, and so on, would 
contribute through diagrams with initial P-wave scattering. As the 
field theory is constructed in this paper, however, the physics of 
these diagrams is implicitly 
included in local short-range operators with growing powers of the photon 
energy $\omega$. Such an operator is explicitly discussed below.

Consequently, there are no additional capture diagrams to consider
at NLO. The only contribution at NLO is due to the change in the
wavefunction renormalization. This leads to a constant factor
\begin{align}
\left(\frac{\sqrt{\mathcal{Z}}}
{\sqrt{\mathcal{Z_{\mrm{LO}}}}}\right)^2=&\frac{1}{\tilde{H}(\gamma,\kC)-3k_\mathrm{C}r_0}\\
=&\frac{\tilde{H}(\gamma,\kC)A^2}{6\kC\Gamma{(1+\kC/\gamma)}^2}
\label{eq:NLOfactor}
\end{align}
larger (or smaller) result than the LO result. It is important to note
that if $r_0\approx\tilde{H}(\gamma,\kC)/(3\kC)$, or equivalently if the ANC $A$ is very
large, then the NLO correction will be large, too. We discuss this in more
detail in
Sec.~\ref{sec:finetuning}.

The next correction enters at N$^4$LO and as such this calculation is valid up to N$^3$LO.

\subsection{N$^4$LO}

\begin{figure}[t]
\centerline{\includegraphics*[scale=0.5,angle=0,clip=true]{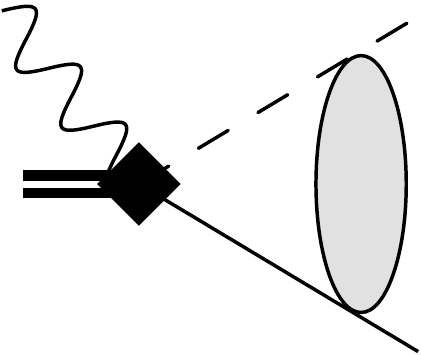}}
\caption{The capture diagram entering at N$^4$LO.}
\label{fig:CaptureN5LO}
\end{figure}

At N$^4$LO a short-range E1 operator appears. The interaction term is
given by
\begin{equation}
\mathcal{L}_\mrm{(N^4LO)}=\left[D_{5/2}^\mrm{(E1)}\mathcal{C}_{is}^a\mathcal{C}_{aj}^{s'}+D_{1/2}^\mrm{(E1)}\mathcal{C}_{is}^\sigma\mathcal{C}_{\sigma j}^{s'}\right]d^\dagger_{s'}\left(\partial_0A_j-\nabla_j A_0\right)\left(\psi_1\overleftrightarrow{\nabla}_i\psi_{0,s}\right)+\mrm{h.c.}~,
\label{eq:shortrangecapture}
\end{equation}
and it has dimension $9$.
This short-range interaction is simply a contact vertex, where the
incoming proton-core pair is in a relative P-wave.

The operator (\ref{eq:shortrangecapture}) gives rise to the capture
diagram in Fig.~\ref{fig:CaptureN5LO}. The tree-level amplitude
is given by
\begin{equation}
\mathcal{B}=D^{(E1)}\sqrt{\mathcal{Z}}\omega\exp{(i\sigma_1)}\pp\sqrt{(1+\eta^2)C_\eta^2}~.
\label{eq:captureN5LO}
\end{equation}
The derivation of the amplitude (\ref{eq:captureN5LO}) involves the
evaluation of the P-wave integral
$\int\frac{\d^3k}{(2\pi)^3}\kk\psi_\pp(\kk)=-\exp{(i\sigma_1)}\pp\sqrt{(1+\eta^2)C_\eta^2}$,
where $\sigma_1$ is the pure Coulomb phase shift in the P-wave (see
App.~B of Ref.~\cite{Ryberg:2014exa} for details on P-wave integrals). The
symbol $D^{(E1)}$ has been introduced as a compact notation for the
total constant of proportionality.

The next relevant operator
\begin{equation}
\mathcal{L}_\mrm{(N^6LO)}=\left[F_{5/2}^\mrm{(E1)}\mathcal{C}_{is}^a\mathcal{C}_{aj}^{s'}+F_{1/2}^\mrm{(E1)}\mathcal{C}_{is}^\sigma\mathcal{C}_{\sigma j}^{s'}\right]d^\dagger_{s'}\partial_0\left(\partial_0A_j-\nabla_j A_0\right)\left(\psi_1\overleftrightarrow{\nabla}_i\psi_{0,s}\right)+\mrm{h.c.}~,
\end{equation}
of dimension $11$, enters at N$^6$LO and as such this
calculation is valid up to N$^5$LO.


\section{Fine tuning and S-wave proton halos}
\label{sec:finetuning}
\begin{figure}[t]
\centerline{\includegraphics*{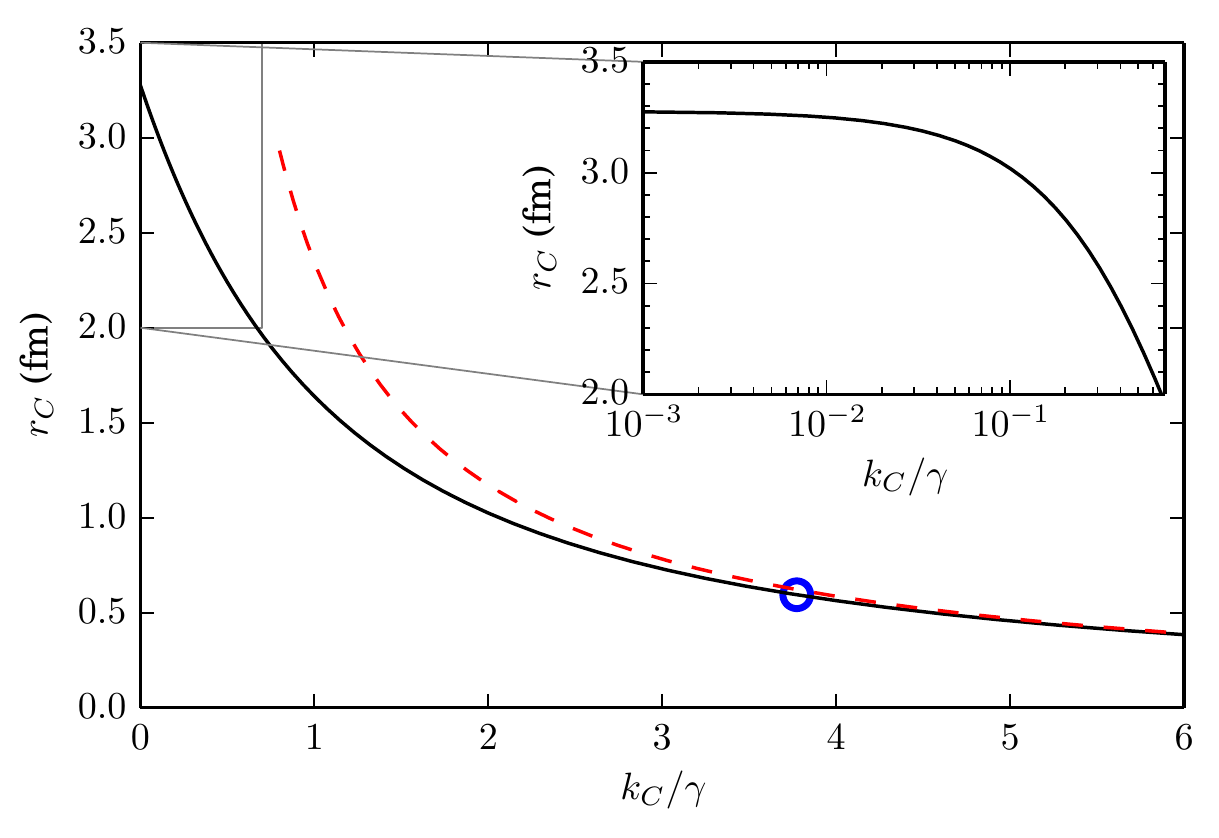}}
\caption{The dependence of the LO charge radius on $\kC/\gamma$. The solid
  black line is the LO Halo EFT result, the blue circle denotes the
  $\nuc{17}{F}^*$ system, and the dashed red
  line is the asymptotic $1/\kC$ behavior. The inset shows the low-energy region in a semi-logarithmic scale, illustrating the hypothetical neutron halo limit~\eqref{eq:rCneutronlimit}. The curve was generated using a
  binding momentum $\gamma=13.6~\mrm{MeV}$.}
\label{fig:rCplotkC}
\end{figure}

Along the neutron drip-line there exist several neutron halo
states. These states are characterized by an unnaturally large
neutron-core scattering length, which brings the state very close to
threshold. However, proton halos are much more rare. In the S-wave
case this can be understood by considering the Coulomb repulsion
between the valence proton and the core. For a proton halo state to
exist, we need the attractive strong force to be almost cancelled by
the Coulomb repulsion, resulting in a threshold 
state~\cite{Zhukov:1995zz,Woods:1997cs}. This
cancellation can be seen within our formalism as an additional
fine-tuning in the effective range. Note that for an S-wave proton
halo state this means that the proton-core scattering length needs to
be unnaturally large and that the effective range must be fine-tuned
to cancel the Coulomb repulsion. The existence of proton halos is
therefore doubly suppressed by the need for two fine-tunings.

Proton halo systems contain the second scale $\kC$ that will depend on
the charge of the degrees of freedom in the system. Within our
framework, $\kC$ is a parameter that is independent of the Coulomb-modified effective range parameters. When $\kC/\gamma\gg1$, we will
speak of the extreme Coulomb regime. In this regime the two particles
tend to be close together, since otherwise the system is ripped apart
by the Coulomb repulsion. This limit is in practice realized for the
$\nuc{17}{F}^*$ system, where $\kC/\gamma=3.8$.  In
Fig.~\ref{fig:rCplotkC} the LO charge radius is shown as a function of
the Sommerfeld parameter $\kC/\gamma$, where we have used the binding momentum for
$\nuc{17}{F}^*$, $\gamma=13.6~\mrm{MeV}$. The blue circle is the parameter point corresponding
to the physical $\nuc{17}{F}^*$ system. It is clear that this system
is almost in the extreme Coulomb regime. Note that the resulting LO
charge radius is very small for a strong Coulomb repulsion, since it
has an asymptotic $1/\kC$ behavior. At the far left, where
$\kC\ll\gamma$, the system mimics that of a neutron halo, with the
only difference being that the photon also can couple through minimal
substitution to the nucleon field. The limiting value for $\kC/\gamma\to 0$ 
is therefore given by \cite{Hammer:2011ye}
\begin{equation}
\lim_{\kC/\gamma\to0}r_\mrm{C}^2=\frac{1}{\Zc+1}\frac{\Zc f^2+(1-f)^2}{2\gamma^2}~.
\label{eq:rCneutronlimit}
\end{equation}

In the standard power counting for systems with large S-wave
scattering length the effective range enters at NLO. The 
hierarchy of scales in this case is $\gamma, \kC \ll 1/r_0\sim 1/R_\mrm{core}$,
where $R_\mrm{core}$ is the length scale set by the core. However, the
discussion above implies that we can have $\kC r_0 \sim 1$ instead
of $\kC \ll 1/r_0$. In the
zero-range limit, the inverse Coulomb momentum sets the scale for 
the LO charge radius and the effective range contributions will
therefore be numerically large since the  
LSZ-factor for S-wave proton halo nuclei with the effective range
included behaves as
\begin{equation}
\mathcal{Z}\propto\frac{1}{1-3\kC r_0}~.
\end{equation}
It appears to be fine tuned to the pole position with
$r_0\sim1/(3\kC)$. In the case of the $\nuc{17}{F}^*$ system the
effective-range correction results in a factor $3.6$--$3.8$ larger
charge radius. 
This hierarchy-of-scales problem can be solved by fixing the bound state pole
position of the t-matrix at leading order and the ANC at NLO. This
procedure ensures that corrections beyond NLO scale naturally again.

\section{S-factor for $^{16}$O$(p,\gamma)^{17}$F$^*$}
\label{sec:results-fluorine-17}

\begin{figure}[t]
\centerline{\includegraphics*[scale=1,angle=0,clip=true]{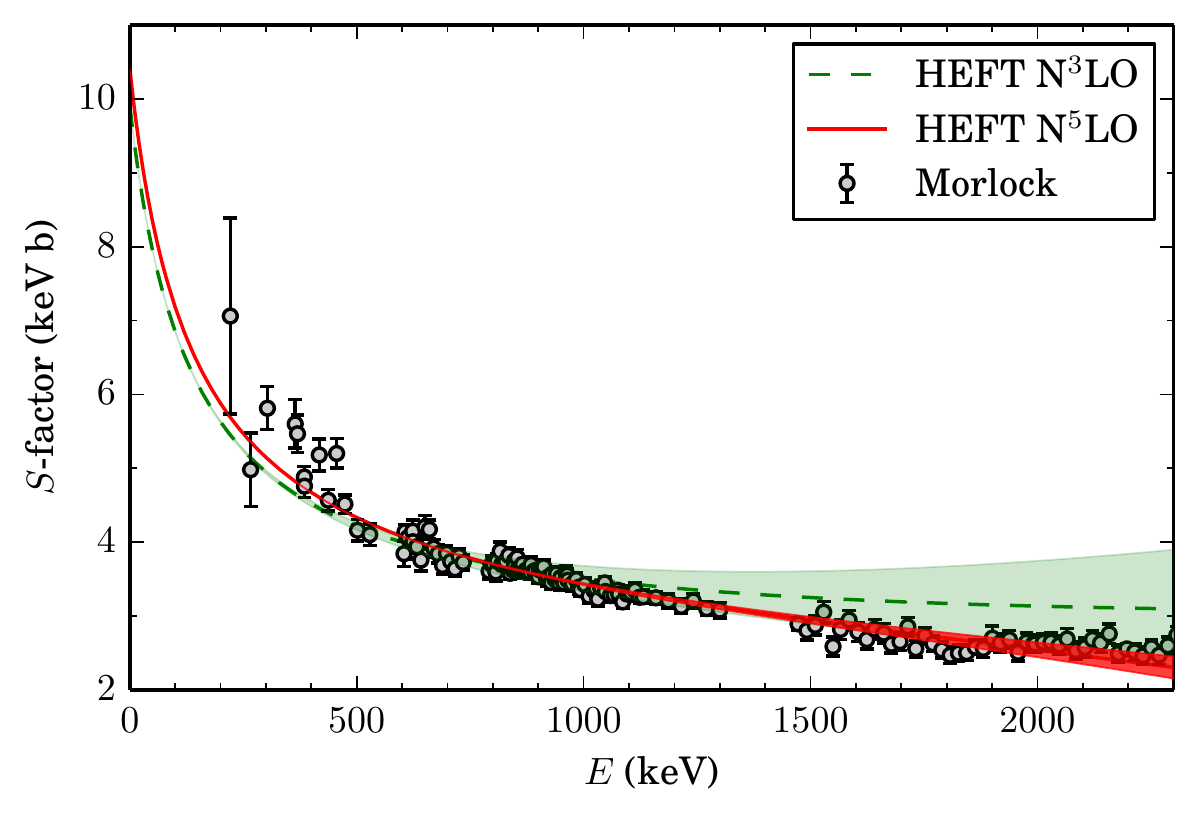}}
\caption{Energy dependent S-factor for $^{16}$O$(p,\gamma)^{17}$F$^*$
fitted to experimental data by
  Morlock {\it et al.}~\cite{Morlock:1997jh}. The error bands correspond to the
  model error from omitted terms at higher orders. See text for details.}
\label{fig:SfactorNLO}
\end{figure}

The excited $1/2^+$ state of $^{17}$F is due to an S-wave
interaction between the valence proton and the \nuc{16}{O}$(0^+)$
core. As such, we describe this proton halo state using the Halo EFT
formalism presented in this paper.

As was discussed in Sec.~\ref{sec:finetuning}, the Coulomb momentum is
larger than the binding momentum for $\nuc{17}{F}^*$, that is
$\kC\gg\gamma$. Consequently, the effective-range correction is needed
for reliable predictions. The ANC for this system has been extracted by Huang {\it et al.} \cite{Huang:2008ye}, using a single-particle model fit of radiative capture data, as $A_1=77.21~\mrm{fm}^{-1/2}$ and experimentally by Gagliardi {\it et al.} \cite{Gagliardi:1998zx}, using the transfer reaction $\nuc{16}{O}(\nuc{3}{He},d)\nuc{17}{F}$, as $A_2=(80.6\pm4.2)~\mrm{fm}^{-1/2}$. Comparing these ANCs
with the LO result (\ref{eq:LOANC}),
$A_\mrm{LO}=21.4~\mrm{fm}^{-1/2}$, makes it clear that the effective
range must be very close to the pole position $1/(3\kC)$, such that
the LSZ residue becomes large.

In Fig.~\ref{fig:SfactorNLO} we present our Halo EFT results for the
radiative proton capture reaction $\nuc{16}{O}(p,\gamma)\nuc{17}{F}^*$
together with data by Morlock {\it et al.} \cite{Morlock:1997jh}. The
green dashed line is the Halo EFT result valid up to N$^3$LO. The
single free parameter of this model is the effective range, and this
was fitted to the experimental data by Morlock 
{\it et al.}~\cite{Morlock:1997jh}
by minimizing an objective function defined as
\begin{align}
  \label{eq:objective_function}
    \chi^2(\vec{\alpha}) \equiv \sum_{i\in \mathbb{M}}
    \left(\frac{\mathcal{O}_i^{\text{th}}(\vec{\alpha}) -
        \mathcal{O}_i^{\text{exp}}}{\delta_i}\right)^2, 
\end{align}
where $\mathcal{O}_i^{\text{th}}$ and $\mathcal{O}_i^{\text{exp}}$
denote the theoretical and experimental values of observable
$\mathcal{O}_i$ in the pool of fit data $\mathbb{M}$, and the total
uncertainty $\delta_i$ determines the weight of the residual. The
theoretical values, and therefore the residuals, depend on the vector
of fit parameters $\vec{\alpha}$. The total uncertainty is the squared
sum of experimental and theoretical errors, $\delta_i^2 =
\delta_{\mrm{exp},i}^2+\delta_{\mrm{th},i}^2$.
The truncation of the Halo EFT expansion allows to estimate the model
error. At N$^3$LO the omitted terms should scale as
$(p/k_\mrm{hi})^4$, where $p$ is the incoming momentum and
$k_\mrm{hi}$ the breakdown scale of the theory. At this order, we
therefore assign a theoretical error
$\delta_{\mrm{th},i}=C_\mrm{EFT}\left(p_i/k_\mrm{hi}\right)^4$.  Taking
these energy-dependent model errors into account will allow us to
include relatively high energy data in the fit.
We use data up to a center-of-mass energy of $2.3~\mrm{MeV}$ and
determine the amplitude, $C_\mrm{EFT}$, of the model error by the
statistical guiding principle that the total $\chi^2$ per degree of
freedom should be unity.  Note that the Morlock data has a
normalization error of $10\%$, which we subtracted in quadrature from
the total experimental error during the fitting procedure. This
normalization error was added back after the fit had been
performed. The constant $C_\mrm{EFT}$ is expected to be of natural
size and is determined iteratively such that the $\chi^2$ per degree
of freedom is minimized to unity. This systematical theory error at
N$^3$LO is shown as a green band in Fig.~\ref{fig:SfactorNLO}, with
$C_\mrm{EFT}=6.9$ and the breakdown scale given by
$k_\mrm{hi}=76~\mrm{MeV}$. Although somewhat large, this value of  
$C_\mrm{EFT}$ is still consistent with our power counting estimate.

Similarly, the red line is given by a fit of the N$^5$LO model to the
same data set where the systematic theory error estimate scales as
$(p/k_\mrm{hi})^6$. At this order we find $C_\mrm{EFT}=1.9$ with
$k_\mrm{hi}=76~\mrm{MeV}$. It is clearly seen that the result
converges with increasing order of the EFT and that the S-factor value
at threshold is stable. From these fits, we extract a threshold
S-factor
\begin{equation}
S=\left\{
\begin{array}{c}
\big(9.9\pm0.1~(\mrm{stat})\pm1.0~\mrm{(norm)})~\mrm{keV~b}~,~\mrm{N^3LO}\\
\big(10.4\pm0.1~(\mrm{stat})\pm1.0~(\mrm{norm})\big)~\mrm{keV~b}~,~\mrm{N^5LO}
\end{array}\right.
\end{equation} 
with the $1\%$ error due to the EFT fit (mainly statistical error) and
the $10\%$ error from the uncertainty in the absolute
normalization of the experimental data. These results give the ANC
\begin{equation}
A=\left\{
\begin{array}{c}
\big(77.4\pm0.2~(\mrm{stat})\pm3.8~(\mrm{norm})\big)~\mrm{fm^{-1/2}}~,~\mrm{N^3LO}\\
\big(79.3\pm0.2~(\mrm{stat})\pm3.9~(\mrm{norm})\big)~\mrm{fm^{-1/2}}~,~\mrm{N^5LO}
\end{array}\right.~,
\end{equation}
which is consistent with the ANCs of Huang {\it et al.} and Gagliardi
{\it et al.}.

The charge radius of the $\nuc{17}{F}^*$ can now be obtained by using
an extracted ANC. Using the $\nuc{16}{O}$-proton ANC extracted from
the N$^5$LO radiative proton capture fit, the resulting NLO charge
radius is given by
\begin{equation}
r_\mrm{C,NLO}=(2.20\pm0.04~(\mrm{EFT})\pm0.11~(\mrm{ANC}))~\mrm{fm}~.
\label{eq:chargeradiusresult}
\end{equation}
The NLO EFT error in Eq.~(\ref{eq:chargeradiusresult}) was estimated
from the EFT expansion parameter squared,
$\left(\gamma/k_\mrm{hi}\right)^2$, using a breakdown scale
$k_\mrm{hi}=76~\mrm{MeV}$, which is of the same order as the inverse
core radius $1/R_\mrm{core}\sim73~\mrm{MeV}$. The dominant error in
Eq.~(\ref{eq:chargeradiusresult}) is from the normalization error of
the Morlock data, through the extracted ANC. However, there could also
be additional EFT errors for this result due to the non-inclusion of the 
operators that are responsible for the finite-size contributions of the 
constituents \cite{rybergbiraforssen}. 
Alternatively, we can interpret our result 
as the radius relative to the $\nuc{16}{O}$ core. 

\section{Conclusions}
\label{sec:conclusions}
We have calculated the charge radius and radiative capture cross
section for proton halo nuclei interacting through a large S-wave
scattering length. Specifically, we have included higher-order
effective-range corrections and shown consequent good agreement with
experimental data. Our description of proton-capture on $^{16}$O
agrees very well with the data and leads to a new way of extracting
the low-energy S-factor and the corresponding ANC with error estimates
that take the intrinsic effective theory error 
into account. For the excited state of $^{17}$F we
found a large correction to all observables at NLO and
have explained this as a result of the complicated interplay between
the range of the strong interaction and the scale set by the Coulomb
interaction. We showed that constraining LO and NLO to proton bound
state pole position and ANC will warrant that higher order corrections
scale naturally. Our results highlight that Halo EFT provides a
powerful tool to analyze experimental data and predict observables 
with theoretical error estimates.
A further advantage of the analysis of such systems with
an EFT framework is the proper identification of two-body
contributions to the electromagnetic current operator. The physics
of such terms should in principle also appear in a cluster model 
description although there are some contributions that are not
generated by gauging the strong interaction terms and are gauge 
invariant by themselves. The EFT
power counting also provides a reliable estimate for the relative 
importance of these terms.

Our analysis can also be used to obtain the low-energy
phaseshifts for nucleon-nucleus elastic scattering that can then be
compared to corresponding \textit{ab initio} calculations that have
become possible over the last years \cite{Hagen:2012rq}.

It is a very interesting question, whether a two-proton halo could in
principle support an effective three-body bound state and how the
spectrum of this system compares to two-neutron halo nuclei bound by
resonant S-wave interactions. This question is also relevant for studies of 
two-proton radioactivity \cite{Grigorenko:2000zz,Blank:2007zz}. 
Furthermore, it seems worthwhile to
investigate the description of low-lying resonances in proton$+$core systems.

\section*{Acknowledgment}
We thank H. Esbensen and S. K\"onig for helpful discussions, and
P. Mohr for supplying relevant data. This work was supported by the
Swedish Research Council (dnr. 2010-4078), the European Research
Council under the European Community’s Seventh Framework Programme
(FP7/2007-2013) / ERC grant agreement no.~240603, the Swedish
Foundation for International Cooperation in Research and Higher
Education (STINT, Grant No.\ IG2012-5158), the Office of Nuclear
Physics, U.S.~Department of Energy under Contract
nos. DE-AC02-06CH11357 and DE-AC05-00OR22725, 
by the BMBF under contracts
05P12PDFTE and 05P15RDFN1, and by the Helmholtz Association under contract
HA216/EMMI.
\begin{appendix}
\section{Partial wave projected Coulomb Green's function}
\label{sec:app}
In this appendix, we discuss the construction and partial wave
projection in the S-wave channel.

It is convenient to use the Coulomb Green's function in a non-integral
form and below we present such a form for the bound state Green's
function. This can be done by doing a partial-wave projection and
forming the Green's function as a product between two independent
Coulomb wavefunctions, satisfying one boundary condition each, in
accordance with the definition of Green's function. For the $r=0$ boundary
condition we must use the regular Coulomb wave function $F_L$ and to
satisfy the condition for a bound state at $r=\infty$ we need to form
the combination
\begin{equation}
iF_L+G_L~.
\label{eq:FGfunction}
\end{equation}
This can be seen from the asymptotics
\begin{equation}
F_L(\eta,\rho)\to\sin{(\rho-L\pi/2-\eta\log{(2\rho)}+\sigma_L)}
\end{equation}
and
\begin{equation}
G_L(\eta,\rho)\to\cos{(\rho-L\pi/2-\eta\log{(2\rho)}+\sigma_L)},
\end{equation}
using that for a bound state $\rho=i\gamma r$, with $\gamma>0$, where
the only combination that yields only an $\exp{(-\gamma r)}$
dependence is the combination given in
(\ref{eq:FGfunction}). Therefore, the partial wave projected Coulomb
Green's function is
\begin{equation}
G^{(L)}_\mathrm{C}(-B;\rho',\rho)=-\frac{m_\mathrm{R}p}{2\pi}\frac{F_L(\eta,\rho')\left[iF_L(\eta,\rho)+G_L(\eta,\rho)\right]}{\rho'\rho}~,
\label{eq:CGF3}
\end{equation}
where the normalization is given by the Wronskian of the Coulomb wavefunctions and the Coulomb-Schr\"odinger equation.

For the S-wave bound state observables we consider, only propagation
with a Coulomb Green's function down to zero separation is needed,
that is $G_\mathrm{C}^{(0)}(-B;0,r)$. Using that
\begin{equation}
iF_L(\eta,\rho)+G_L(\eta,\rho)=\sqrt{\frac{\Gamma{(L+1+i\eta)}}{\Gamma{(L+1-i\eta)}}}
\exp{\left(-i\frac{\pi}{2}(L+i\eta)\right)}W_{-i\eta,L+1/2}(-2i\rho)~,
\end{equation}
which was shown in \cite{Koenig:2012bv}, and the limit
\begin{equation}
\lim_{\rho\to0}\frac{F_0(\eta,\rho)}{\rho}=\exp{(-\pi\eta/2)}
\sqrt{\Gamma{(1+i\eta)}\Gamma{(1-i\eta)}}
\end{equation}
we may write Eq.~(\ref{eq:CGF3}) for S-waves as
\begin{equation}
G^{(0)}_\mathrm{C}(-B;0,\rho)=-\frac{m_\mathrm{R}p}{2\pi}\Gamma{(1+i\eta)}
\frac{W_{-i\eta,1/2}(-2i\rho)}{\rho}~.
\label{eq:CGF4}
\end{equation}

\end{appendix}

\end{document}